\documentclass[11pt,a4paper]{article}

\usepackage[utf8]{inputenc}
\usepackage[T1]{fontenc}
\usepackage[english]{babel}

\usepackage{amsmath,amssymb}
\usepackage{geometry}
\usepackage{graphicx}
\usepackage{caption}
\usepackage{subcaption}
\usepackage{hyperref}
\usepackage{float}

\numberwithin{equation}{section}

\title{Magnetic Phase Control of a Thick SNS Weak Link:\\[4pt]
Proposed experimental scheme}

\author{Aleksey Turchanov}

\date{November 2025}

\begin{document}

\maketitle

\begin{center}
\small
\scriptsize
Licensed under Creative Commons Attribution 4.0 International (CC BY 4.0).\\
Zenodo record: DOI\,\texttt{10.5281/zenodo.17703630}.
\end{center}

\vspace{1em}

\begin{abstract}

\vspace{0.8em}
In contrast to the extensive literature on thin tunnel junctions and traditional SQUID geometries, there is almost no quantitative experimental data on magnetic control of the Josephson phase in thick SNS weak links. The standard view is that in such compact structures without macroscopic loops the local magnetic coupling to the phase is negligibly small, which in practice forces one to use bulky SQUID devices for phase control.
\\

We show that this view is overly restrictive. The basic idea is simple: a local microcoil is placed directly above a thick SNS bridge and controls the Josephson phase via strong phase--flux coupling, enhanced near the Josephson plasma resonance. In the proposed configuration, realistic thick SNS weak links (with normal-layer thickness $d \sim \xi$) controlled by an on-chip microcoil can reach phase--flux efficiencies of order $30$--$60\,\%$ of an ideal dc SQUID.
\\

Within a standard RSJ/RCSJ model and linear circuit theory, we show that such unexpectedly strong coupling arises from the combination of a large kinetic inductance of the thick SNS bridge and resonant amplification of about $15$--$35$ dB, rather than from any exotic microphysics. The proposed experiment---a comparative analysis of Shapiro steps driven by a direct RF signal and by the magnetic field of the same microcoil---provides a direct and quantitative method to measure the phase--flux response of a thick SNS junction.
\\

If confirmed experimentally, such structures may become the first compact ``phase elements'' capable of locally controlling the Josephson phase without a macroscopic loop, opening the way to a radical simplification of superconducting quantum circuit architectures with dense, locally addressable phase control.
\end{abstract}

\newpage

%======================================================================
\section{Introduction}
%======================================================================

Superconducting electronics relies critically on precise control of the Josephson phase in weak links between superconductors. In practice, magnetic control of this phase is almost always implemented via macroscopic SQUID loops, which limits integration density and complicates circuit architectures. It is widely assumed that compact structures without large loops cannot provide sufficiently strong magnetic control of the Josephson phase to be of practical use.

In this work we challenge this assumption on the example of thick SNS (superconductor--normal metal--superconductor) weak links, where the thickness of the normal layer is comparable to the superconducting coherence length. Our central question is: can a local on-chip microcoil, placed directly above such a thick SNS bridge, provide efficient magnetic control of the Josephson phase, and if so, under what conditions?

We address this question entirely within a standard classical framework: conventional Josephson relations in the RSJ/RCSJ formulation combined with linear circuit theory, without introducing any modified or exotic microphysics. From the circuit point of view, the device we consider is a Josephson oscillator (the thick SNS weak link) driven by a local magnetic source (the microcoil). We show that realistic thick SNS bridges with an integrated microcoil can, in principle, reach phase--flux control efficiencies on the order of $30$--$60\%$ of an ideal dc SQUID, while requiring no macroscopic loop.

Within this classical RSJ/RCSJ framework we explain why thick SNS structures can exhibit such unexpectedly strong coupling: the combination of a large kinetic inductance of the thick bridge, efficient mutual coupling to the on-chip microcoil, and resonant enhancement near the Josephson plasma frequency leads to a strong, frequency-dependent link between the local magnetic field and the phase. We quantitatively evaluate the efficiency of this phase--flux conversion and identify ranges of device parameters (normal-layer thickness, inductances, quality factor) for which local magnetic phase control becomes practically significant. The paper provides a convenient calculation scheme that, for a given device geometry and drive frequency, yields the expected phase response and thus allows such elements to be designed in a targeted way.

For experimental verification we propose a concrete protocol based on a comparative analysis of Shapiro steps generated by (i) direct RF excitation and (ii) magnetic excitation by the same microcoil. This approach does not require detailed microscopic modeling of the SNS region and directly outputs a measurable figure of merit for phase control efficiency in a real device. The purpose of the present paper is therefore twofold: (i) to derive transparent
classical estimates for the strength of local magnetic phase control in thick SNS
weak links, and (ii) to formulate a minimal, theory-driven protocol for testing
these estimates. Section~2 describes the device geometry and the coarse-grained
phase-sensitive region. Section~3 presents the classical phase--flux model and
order-of-magnitude estimates; Sections~4--6 discuss the proposed experimental
configuration, possible outcomes, and their implications for the design of dense
superconducting circuits with locally addressable phase control.

\newpage
%======================================================================
\section{Device and coarse-grained description}
\label{sec:device}
%======================================================================

\subsection{Device geometry and field-active region}

\begin{figure}[H]
  \centering
  \includegraphics[width=0.95\linewidth]{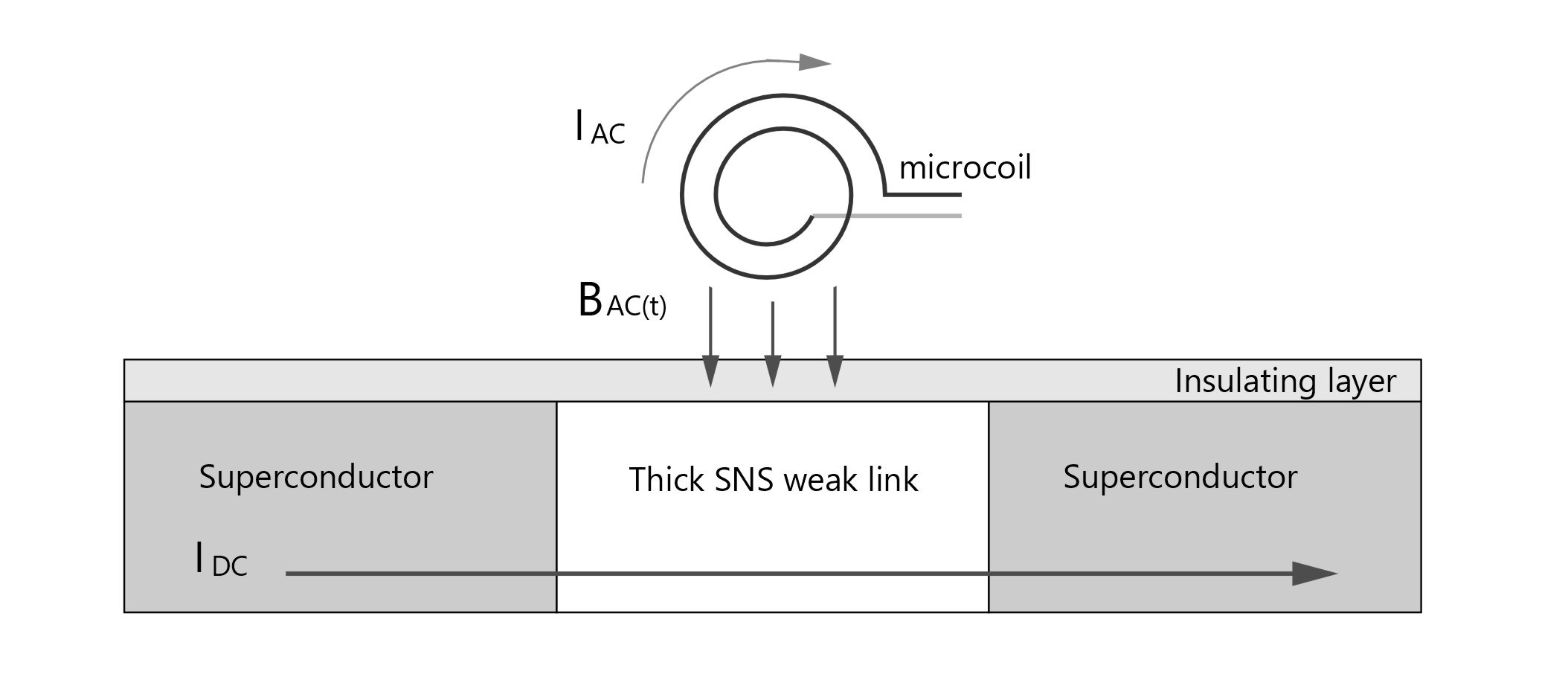}
  \caption{Cross-sectional sketch of the proposed device (not to scale). A superconducting
  strip is interrupted by a thick SNS weak link of length $L_W$ and thickness $d$. An
  on-chip microcoil above the insulating layer carries an AC current $I_{\mathrm{AC}}$,
  generating a local AC magnetic field $B_{\mathrm{AC}}(t)$ through the weak link. A DC
  bias current $I_{\mathrm{DC}}$ flows along the superconducting strip. This figure defines
  the geometric parameters used in the classical Josephson--circuit model.}
  \label{fig:device}
\end{figure}

The proposed device consists of a superconducting strip interrupted by a thick SNS weak
link of length $L_W\approx 500~\mathrm{nm}$ and thickness
$d\approx 20$--$40~\mathrm{nm}$. An on-chip microcoil is fabricated above an insulating
layer on top of the weak link and carries an AC current $I_{\mathrm{AC}}(t)$, generating a
local AC magnetic field $B_{\mathrm{AC}}(t)$ through the weak link, as shown in
Fig.~\ref{fig:device}. A DC bias current $I_{\mathrm{DC}}$ flows along the strip.

\begin{figure}[H]
  \centering
  \includegraphics[width=0.95\linewidth]{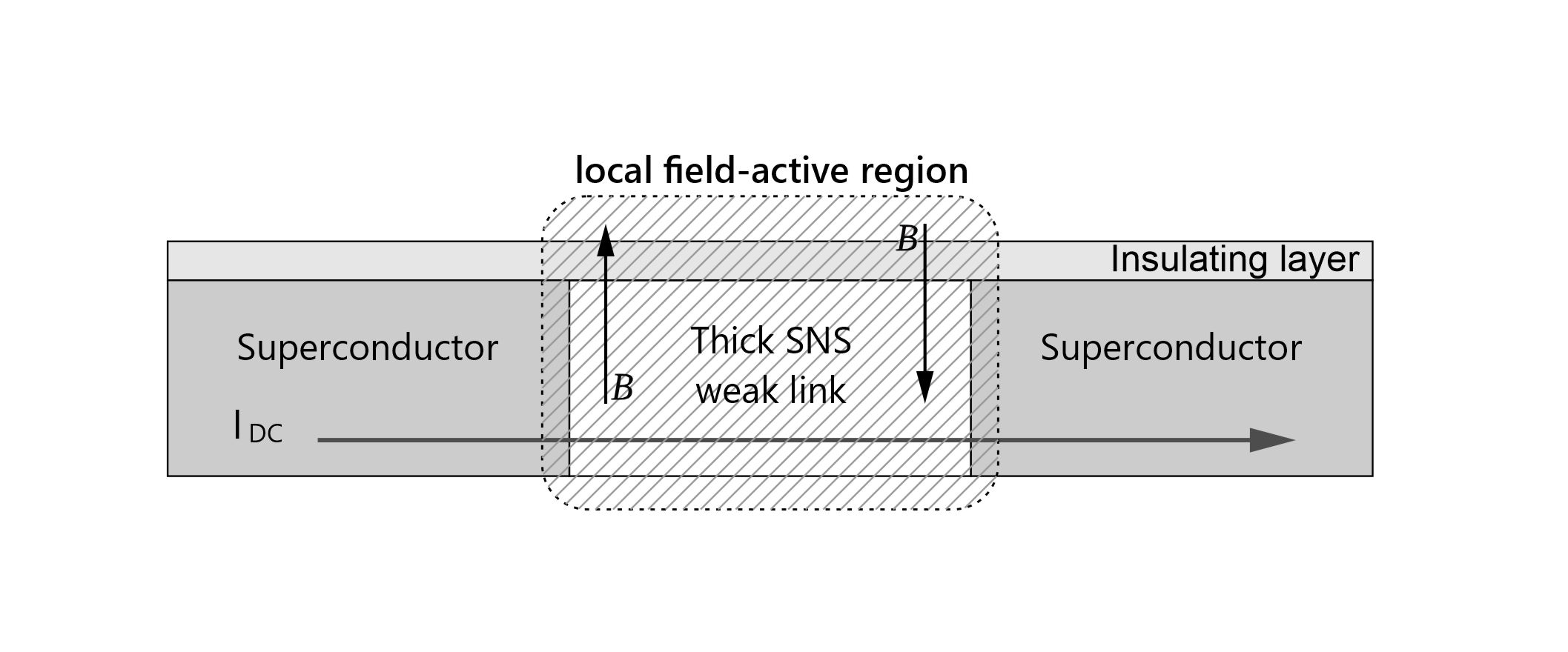}
  \caption{Same geometry as Fig.~\ref{fig:device}, with the hatched, dashed region
  indicating the local field-active volume where the microcoil field penetrates the thick
  SNS weak link and neighbouring superconducting material. In the classical description
  this is the effective phase-sensitive region that determines the flux $\Phi_{\mathrm{ext}}(t)$
  entering the phase--flux coefficient $\alpha(d,\omega)$.}
  \label{fig:field_active}
\end{figure}

It is important to emphasise that the proposed device probes the field-active
region of Fig.~\ref{fig:field_active} only through its net phase response. Any microscopic
Andreev-type states or vortex textures that may exist inside the thick SNS weak link are
not resolved individually. Their combined effect is absorbed into the effective inductance
and into the phase--flux coefficient $\alpha(d,\omega)$ discussed below and compared to the
experimental value $\alpha_{\mathrm{exp}}(d,\omega)$ extracted from the Shapiro-step
protocol in Sec.~\ref{sec:experiment}.

For concreteness, one may think of a Nb/Cu/Nb bridge patterned on a Si/SiO$_2$
substrate, with a normal-layer thickness $d$ in the 20--40~nm range and a strip width of a few hundred nanometres, as in the order-of-magnitude estimates of Sec.~\ref{sec:oom_estimate}. However, the analysis in this section and the protocol
described in Sec.~\ref{sec:experiment} do not rely on any particular material choice: different laboratories can
implement essentially the same geometry with their preferred superconducting and
normal-metal stacks, as long as the resulting kinetic inductance $L_{\mathrm{kin}}$, mutual
inductance $M$ and quality factor $Q$ fall in the broad parameter ranges discussed in
the classical model.

In thick SNS bridges the microscopic current and phase distribution can be complex: vortex-like currents, proximity-induced Andreev structures, and non-uniform supercurrent flow are all possible. In the present work we deliberately do not resolve these microscopic details. Instead, the entire weak-link region together with the adjacent superconducting banks is replaced by a single effective classical circuit characterised by  
\begin{itemize}
  \item kinetic inductance $L_{\text{kin}}$,
  \item capacitance $C$,
  \item mutual inductance $M$ to the on-chip microcoil,
  \item quality factor $Q$ of the plasma mode.
\end{itemize}

All microscopic physics (Andreev bound states, vortex motion, etc.) enters the calculation exclusively through the numerical values of these effective lumped-element parameters — exactly as is standard practice in the conventional RCSJ description of any Josephson weak link.

\newpage
%======================================================================
\section{Classical phase--flux theory}
\label{sec:theory}
%======================================================================

\subsection{RCSJ dynamics and definition of the phase--flux coefficient}
\label{sec:rcsj}

We describe the thick SNS weak link by a single Josephson phase $\varphi(t)$ obeying
the standard RCSJ equation
\begin{equation}
  C \ddot{\varphi} + \frac{1}{R}\dot{\varphi} + I_c \sin\varphi
    = I_{\text{bias}}(t),
  \label{eq:rcsj}
\end{equation}
where $C$ is the effective capacitance, $R$ the shunt resistance,
$I_c$ the critical current and $I_{\text{bias}}(t)$ the applied bias current. The voltage
across the weak link is given by the Josephson relation
\begin{equation}
  V(t) = \frac{\Phi_0}{2\pi}\,\dot{\varphi}(t),
  \qquad
  \Phi_0 = \frac{h}{2e}.
  \label{eq:josephson_relation}
\end{equation}

A current $I_{\text{coil}}(t)$ flowing in the on-chip microcoil produces an effective
magnetic flux threading the kinetic inductance of the bridge,
\begin{equation}
  \Phi_{\text{eff}}(t) = M I_{\text{coil}}(t),
  \label{eq:phi_eff}
\end{equation}
where $M$ is the mutual inductance between the coil and the effective phase-sensitive
region of the weak link.

From the point of view of the RCSJ dynamics this flux enters only through a small
additional phase drive, so that in the linear-response regime we can write
\begin{equation}
  \varphi(t) = \varphi_0(t) + \delta\varphi_{\text{coil}}(t),
  \label{eq:phi_total}
\end{equation}
with $\varphi_0(t)$ the phase evolution in the absence of microcoil drive.

It is convenient to parametrise the small-signal phase response to $\Phi_{\text{eff}}$ by a
dimensionless phase--flux coefficient $\alpha(d,\omega)$ defined by
\begin{equation}
  \frac{\partial\varphi}{\partial\Phi_{\text{eff}}}\bigg|_{\text{small signal}}
  = \frac{2\pi}{\Phi_0}\,\alpha(d,\omega).
  \label{eq:alpha_def}
\end{equation}
In this normalisation an ideal dc SQUID with negligible loop inductance has
$\alpha = 1$, so that values $|\alpha|\sim 0.3$--$0.6$ correspond to 30--60\% of the ideal
dc SQUID phase--flux efficiency.

\subsection{Coarse-grained circuit model for a thick SNS bridge}
\label{sec:coarse_grained}

As discussed above, our aim is not to resolve microscopic Andreev or vortex
configurations, but to describe the integrated phase response of the thick SNS region
under local coil drive. For this purpose we replace the detailed GL+Maxwell problem by the same
effective lumped-element circuit as in Sec.~2: the weak-link region plus its
immediate superconducting surroundings are represented by a kinetic inductance
$L_{\text{kin}}$, a capacitance $C$, a mutual inductance $M$ to the on-chip
microcoil, and a plasma mode with frequency $\omega_p$ and quality factor $Q$.

In this picture the coil field drives the phase of the effective GL+Maxwell mode, and
the phase--flux coefficient $\alpha(d,\omega)$ simply measures how efficiently this
mode converts coil flux into Josephson phase. The hatched region in
Fig.~\ref{fig:field_active}---the ``local field-active volume''---is precisely the spatial
support of this coarse-grained mode. Microscopic Andreev states and vortex textures
modify the response only through the effective parameters $L_{\text{kin}}$, $M$ and $Q$.

At low frequencies, where inductive screening dominates, the static phase--flux
response is controlled by the ratio $M/L_{\text{kin}}$. We capture this by a static
coefficient
\begin{equation}
  \alpha_{\text{static}} \simeq \frac{M}{L_{\text{kin}}},
  \label{eq:alpha_static_basic}
\end{equation}
which can already be orders of magnitude larger for a thick SNS bridge than for a thin
tunnel junction, because the effective field-active volume and kinetic inductance of the
bridge are much larger.

\subsection{Frequency dependence and resonant enhancement of phase response}
\label{sec:freq_dependence}

We now consider the frequency dependence of the phase--flux coefficient. In the
linearised RCSJ model the small-signal dynamics of the phase deviation
$\delta\varphi(t)$ driven by an AC coil current $I_{\text{coil}}(t)=I_0\cos\omega t$ can be
cast in the form of a driven damped harmonic oscillator,
\begin{equation}
  \ddot{\delta\varphi}
  + \frac{\omega_p}{Q}\,\dot{\delta\varphi}
  + \omega_p^2\,\delta\varphi
  = \frac{2\pi}{\Phi_0}\,\frac{M}{L_{\text{kin}}}\,\omega_p^2 I_0\cos\omega t,
  \label{eq:lin_osc}
\end{equation}
where $\omega_p$ is the plasma frequency associated with the kinetic inductance and
capacitance of the bridge and $Q$ is the quality factor of this mode. Solving for the
steady-state response at frequency $\omega$ yields the following expression for the
magnitude of the phase--flux coefficient:
\begin{equation}
  \bigl|\alpha(d,\omega)\bigr|
  = \frac{\alpha_{\text{static}}}{\sqrt{
      \bigl(1-\omega^2/\omega_p^2\bigr)^2
      + (\omega/\omega_p Q)^2 }}.
  \label{eq:alpha_eff_omega}
\end{equation}

In the quasistatic regime $\omega\ll\omega_p$ one recovers the static value
$\bigl|\alpha(d,\omega)\bigr|\approx \alpha_{\text{static}}$. Near the plasma resonance,
$\omega\simeq\omega_p$, the response is enhanced by approximately the quality factor:
\begin{equation}
  \alpha_{\text{res}} \equiv \bigl|\alpha(d,\omega_p)\bigr|
  \simeq \alpha_{\text{static}} Q.
  \label{eq:alpha_res_Q}
\end{equation}
Thus even a modest static coupling $|\alpha_{\text{static}}|\sim 0.03$--$0.07$ can be
amplified into a strong-coupling regime $|\alpha_{\text{res}}|\sim 0.3$--$0.6$ if the plasma
mode has a realistic quality factor $Q\sim 15$--$35$.

\subsection{Order-of-magnitude estimate for a thick SNS bridge with microcoil}
\label{sec:oom_estimate}

We now give a back-of-the-envelope classical estimate for $\alpha(d,\omega)$ in a
representative thick SNS weak link with a planar microcoil placed above it. The goal
is not to replace full GL/Usadel + London--Maxwell simulations, but to show that the
strong-coupling regime $|\alpha|\sim 0.3$--$0.6$ is compatible with realistic device
parameters.

Consider a diffusive SNS weak link with
\begin{itemize}
  \item normal-layer thickness $d \sim 20$--$40~\mathrm{nm}$,
  \item width $w \sim 300~\mathrm{nm}$,
  \item coherence length $\xi$ such that $d \sim \xi$ (thick SNS regime),
  \item London penetration depth in the electrodes $\lambda_L \sim 250~\mathrm{nm}$.
\end{itemize}
For such dimensions, the effective kinetic inductance of the weak-link region can be of
order
\begin{equation}
  L_{\mathrm{kin}} \sim 20\text{--}30~\mathrm{pH},
  \label{eq:Lkin_est}
\end{equation}
while realistic planar microcoils positioned a few hundred nanometres above the bridge
can reach mutual inductances in the range
\begin{equation}
  M \sim 0.8\text{--}1.8~\mathrm{pH}.
  \label{eq:M_est}
\end{equation}
In a simple lumped picture, the static phase--flux coupling can then be estimated as
\begin{equation}
  \alpha_{\mathrm{static}} \sim \frac{M}{L_{\mathrm{kin}}}
  \sim 0.03\text{--}0.07,
  \label{eq:alpha_static_M_Lkin}
\end{equation}
already two--three orders of magnitude larger than the tunnel-junction estimates
$|\alpha|\sim 10^{-3}$--$10^{-2}$.

For localised plasma-like modes in the 5--20~GHz range, quality factors of order
\begin{equation}
  Q \sim 15\text{--}35
  \label{eq:Q_est}
\end{equation}
are not unreasonable for carefully engineered on-chip structures. Combining
Eqs.~\eqref{eq:alpha_res_Q}, \eqref{eq:alpha_static_M_Lkin} and \eqref{eq:Q_est} yields a
resonant phase--flux coupling
\begin{equation}
  |\alpha_{\mathrm{res}}|
  \sim |\alpha_{\mathrm{static}}|\,Q
  \sim 0.3\text{--}0.6.
  \label{eq:alpha_res_window}
\end{equation}

This estimate shows that thick SNS bridges with realistic microcoils can, in
principle, reach a strong-coupling regime in which local magnetic fields produce
sizeable phase modulations without any macroscopic loop. The experimental proposal in
Sec.~\ref{sec:experiment} is designed to test whether this regime is actually realised
in fabricated devices by extracting an operational value $\alpha_{\mathrm{exp}}(d,\omega)$
from Shapiro-step measurements.

Within this classical estimate, the dependence of the coupling strength on the ratio
$d/\xi$ between the normal-layer thickness and the coherence length can be summarised
as follows. Thin tunnel junctions with $d\ll\xi$ exhibit weak magnetic coupling because
the effective field-active volume is small and strongly screened by the electrodes. Overly
thick weak links with $d\gg\xi$ also yield weak coupling, since the Josephson current and
associated phase stiffness are strongly suppressed in the normal region. In contrast, a
thick SNS bridge with $d\sim\xi$ can support a robust proximity-induced supercurrent
over a relatively large volume that is efficiently threaded by the microcoil field, leading
to a strong-coupling regime with $|\alpha|\sim 0.3$--$0.6$.

\begin{figure}[H]
  \centering
  \includegraphics[width=0.75\linewidth]{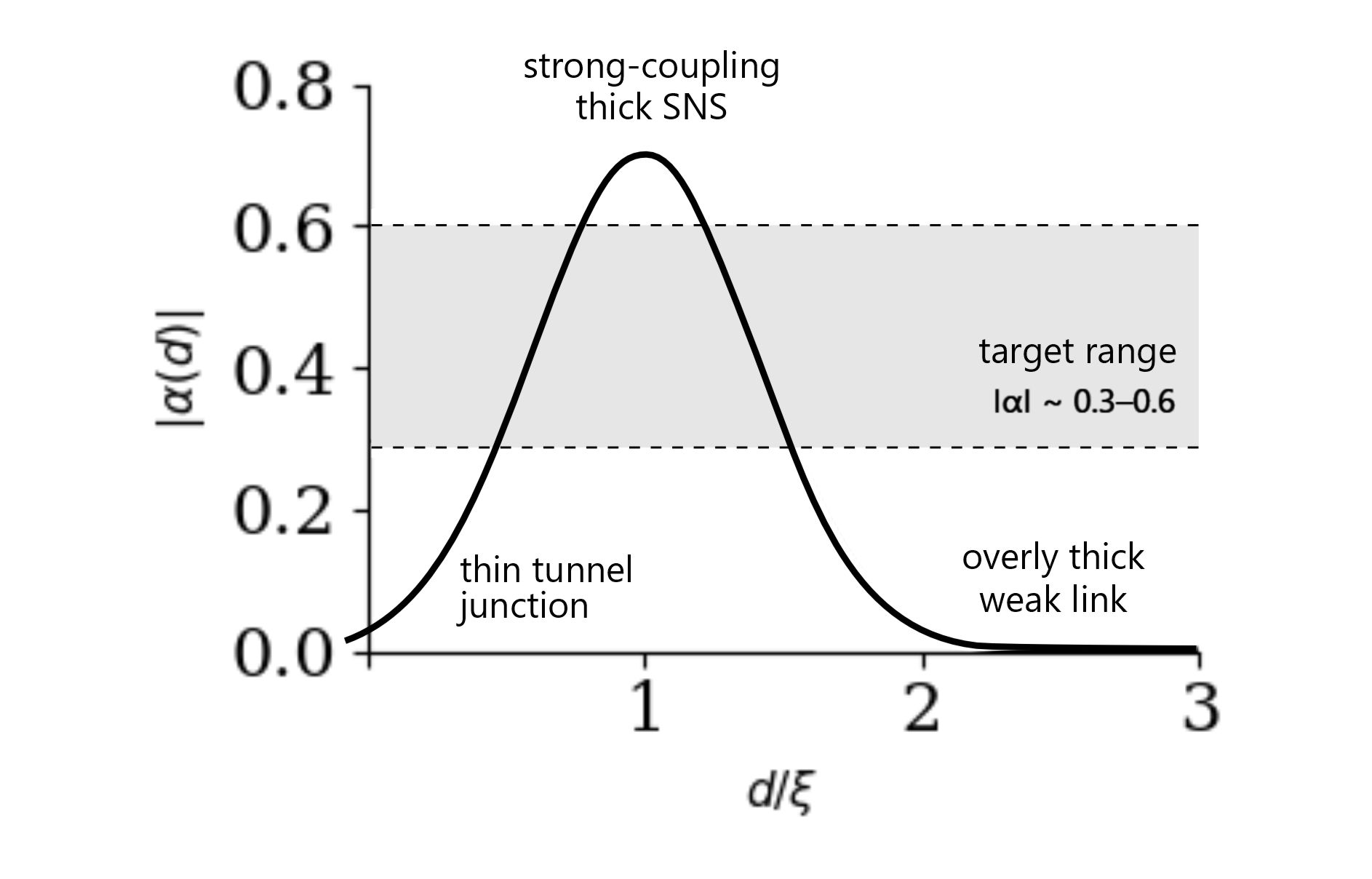}
  \caption{Illustrative dependence of the magnitude of the static phase--flux coefficient
  $|\alpha_{\text{static}}(d)|$ on the ratio $d/\xi$ between the normal-layer thickness
  and the coherence length, within the classical model of Sec.~\ref{sec:theory}. Thin tunnel junctions
  ($d\ll\xi$) and overly thick weak links ($d\gg\xi$) exhibit weak coupling, while a thick
  SNS bridge with $d\sim\xi$ can reach a strong-coupling regime with
  $|\alpha_{\text{static}}|\sim 0.3$--$0.6$, indicated by the shaded band.}
  \label{fig:alpha_d}
\end{figure}

\begin{figure}[H]
  \centering
  \includegraphics[width=0.75\linewidth]{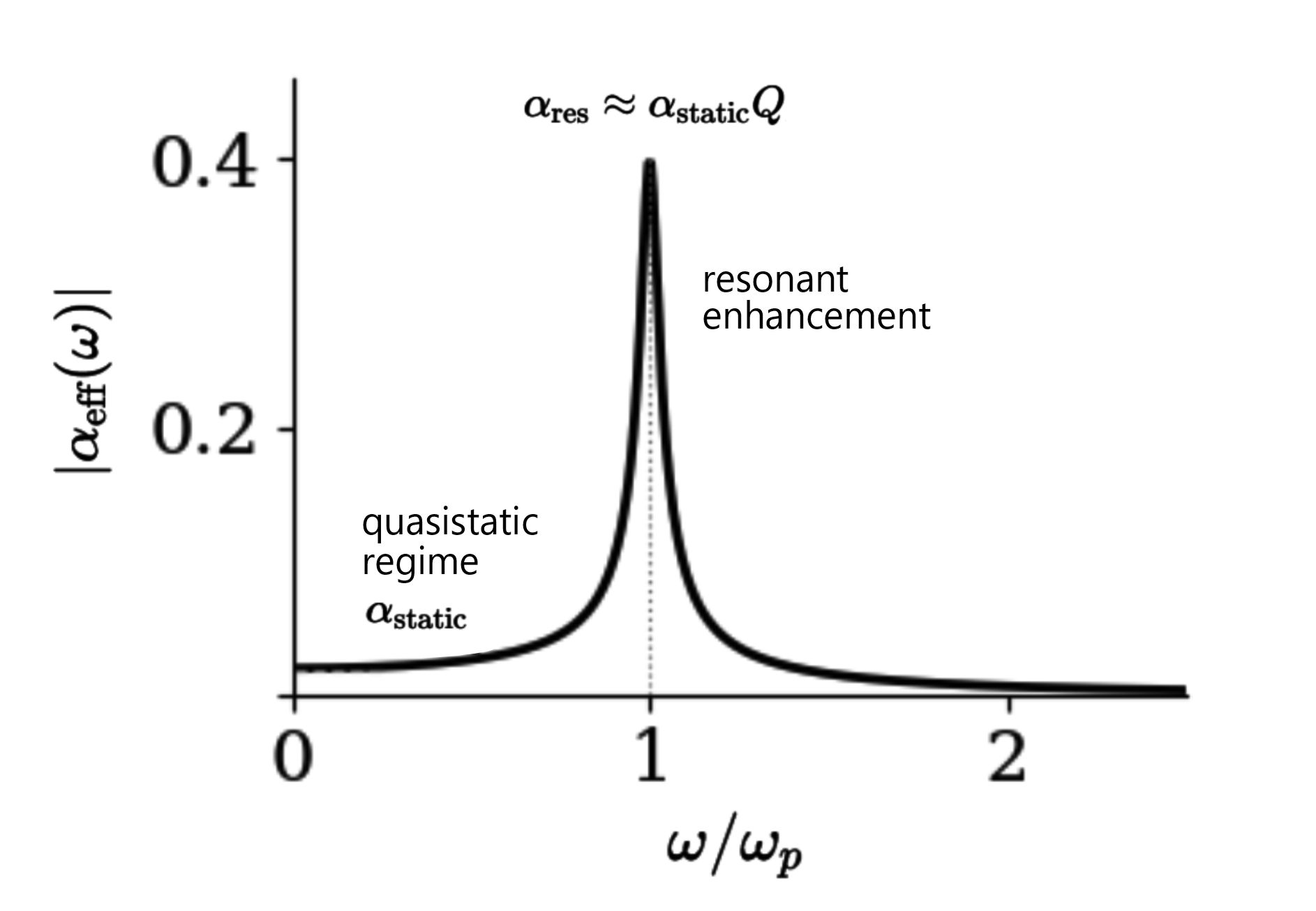}
  \caption{Frequency dependence of the magnitude of the phase--flux coefficient
  $|\alpha(d,\omega)|$ obtained from the linearised RCSJ model. At low
  frequencies $\omega\ll\omega_p$ one recovers the quasistatic value
  $\alpha_{\mathrm{static}}$, while near the plasma resonance $\omega\simeq\omega_p$ the
  coupling is enhanced to $\alpha_{\mathrm{res}}\approx \alpha_{\mathrm{static}} Q$ in
  agreement with Eq.~\eqref{eq:alpha_res_Q}, with $Q$ the quality factor of the mode.}
  \label{fig:alpha_omega}
\end{figure}

\newpage

%======================================================================
\section{Proposed experimental scheme}
\label{sec:experiment}
The aim of this section is not to provide a full experimental design, but to outline a minimal protocol that allows the classical phase--flux coefficient $\alpha(d,\omega)$ of Sec.~\ref{sec:theory} to be tested and quoted as an operational number $\alpha_{\mathrm{exp}}(d,\omega)$ in a real device.

%======================================================================

\subsection{Measurement setup and bias protocol}

In practice, the device would be mounted in a cryogenic environment (for example, a
$^4$He or dilution refrigerator) at temperatures well below the critical temperature of the
superconducting electrodes. The weak link is measured in a standard four-terminal
configuration: a low-noise current source provides the DC bias $I_{\mathrm{DC}}$ through the
strip, and the resulting voltage across the SNS region is recorded by a differential
amplifier and digitiser. The DC lines are filtered and thermalised at intermediate stages
to suppress external noise.

Microwave excitation is delivered either by an on-chip transmission line or by a nearby
antenna, and by the on-chip microcoil itself. In both cases the drive at frequency
$\omega$ originates from a room-temperature microwave source, with appropriate
attenuation and, if necessary, additional cold attenuation close to the sample. The same
biasing and readout setup is used for both parts of the protocol: first with direct RF
drive, then with purely magnetic drive through the coil. Specific details of the wiring,
filters and amplifiers are laboratory-dependent, but from the point of view of the present
proposal they enter only through the effective quality factor $Q$ and the calibration of
the drive amplitudes.

\subsection{Calibration with RF drive}

As a first step, the junction is driven by a conventional RF signal applied through an
on-chip line or a nearby antenna. Within the classical RSJ model this produces a phase
modulation
\begin{equation}
  \varphi(t) = \varphi_{\text{bias}}(t) + a_{\mathrm{RF}}\sin\omega t,
  \label{eq:rf_drive}
\end{equation}
where $a_{\mathrm{RF}}$ is the dimensionless phase amplitude proportional to the RF drive
strength. The resulting Shapiro steps in the $I$--$V$ characteristics have heights governed
by Bessel functions $J_n(a_{\mathrm{RF}})$, providing a direct way to calibrate
$a_{\mathrm{RF}}$ as a function of the applied RF power~\cite{Shapiro1963,Tinkham}.

\subsection{Magnetic drive with on-chip microcoil}

In the second step, the microcoil is used as the sole drive at the same frequency, with
the RF line turned off. The coil current $I_{\mathrm{AC}}$ induces a phase modulation
$a_{\mathrm{mag}}$ through Eq.~\eqref{eq:alpha_def}. The Shapiro steps observed under
purely magnetic drive can then be compared to the RF-calibrated case. By matching the
visible Shapiro plateaus, one extracts an experimental value $\alpha_{\mathrm{exp}}(d,\omega)$
that relates $a_{\mathrm{mag}}$ to $I_{\mathrm{AC}}$. We use $\alpha_{\mathrm{exp}}(d,\omega)$ as an
operational phase--flux coefficient for a given device at drive frequency $\omega$, to be
compared with the theoretical $\alpha(d,\omega)$ obtained in Sec.~\ref{sec:theory}.

\subsection{Expected experimental uncertainties}

The proposed protocol is designed to yield $\alpha_{\mathrm{exp}}(d,\omega)$ as a
well-defined experimental number, while keeping the discussion of laboratory-specific
details minimal. The dominant sources of uncertainty can be summarised as follows.

First, the calibration of the RF-driven phase amplitude $a_{\mathrm{RF}}$ is limited by the
accuracy with which the RF power at the sample is known. Reflections and frequency
dependent losses in the microwave line introduce a systematic uncertainty that
propagates into the inferred $a_{\mathrm{RF}}$ from the Shapiro-step pattern. Second, the
conversion between coil current $I_{\mathrm{AC}}$ and effective flux $\Phi_{\mathrm{ext}}$ depends on
the mutual inductance $M$, which in turn is set by the exact geometry and alignment of
the microcoil relative to the bridge. Small deviations from the nominal spacing and
lateral positioning translate into an uncertainty in $M$.

Finally, residual voltage noise and slow drifts in the DC bias or temperature affect the
precision with which the step heights and plateau structure can be resolved. In a typical
set-up these effects can be reduced but not completely eliminated. As a result, the
extracted $\alpha_{\mathrm{exp}}(d,\omega)$ will carry an experiment-specific error bar whose
size is determined by the particular laboratory implementation.

From the standpoint of the present work, it is sufficient that the overall uncertainty in
$\alpha_{\mathrm{exp}}$ remains small compared to the wide separation between the
conventional weak-coupling regime $|\alpha|\ll 10^{-2}$ and the strong-coupling window
$|\alpha|\sim 0.3$--$0.6$. Fine-grained optimisation of the measurement chain and error
budget would naturally be carried out by individual experimental groups in the course of
a concrete implementation.

%======================================================================
\section{Possible outcomes}
%======================================================================

The proposed experiment is designed to deliver a single quantitative number, the
effective phase--flux coefficient $\alpha_{\mathrm{exp}}(d,\omega)$, extracted from the
comparison between RF-driven and magnetically driven Shapiro steps in a thick SNS
weak link.

If the analysis yields $|\alpha_{\mathrm{exp}}|\ll 10^{-2}$, this would strongly support the
conventional view that local magnetic control of the Josephson phase in small, loopless
junctions is negligible. In this case, thick SNS weak links with on-chip microcoils behave,
for practical purposes, as ordinary weak links whose phase can only be efficiently
controlled via macroscopic flux loops, and the strong-coupling window discussed above
is not realised in the fabricated devices.

If, on the other hand, $\alpha_{\mathrm{exp}}(d,\omega)$ is found to lie within the
strong-coupling window $|\alpha_{\mathrm{exp}}|\sim 0.3$--$0.6$, the result would
demonstrate that a thick SNS weak link in a realistic microcoil geometry can function as
a strongly coupled magnetic phase element without any macroscopic loop. This would
identify thick SNS structures as promising building blocks for locally addressable phase
control in superconducting circuits.

Intermediate values $10^{-2}\lesssim|\alpha_{\mathrm{exp}}|\lesssim 0.1$ would indicate a
partially enhanced, but still moderately weak, magnetic-to-phase coupling. Such outcomes
would suggest that some ingredients of the classical model --- such as the assumed
quality factor, effective kinetic inductance or field distribution --- are only partially
realised in the devices, and would motivate refining both the microscopic modelling and
the device design.

From this point of view, the extracted value $\alpha_{\mathrm{exp}}(d,\omega)$ should be
interpreted as an integral measure of the phase sensitivity of the coarse-grained
GL+Maxwell mode associated with the thick SNS region, rather than as a probe of any
specific Andreev state or individual vortex~\cite{Likharev1979,Pannetier2000}.

%======================================================================
\section{Discussion}
\label{sec:discussion}
%======================================================================

From the standpoint of standard Josephson physics~\cite{Tinkham,BaronePaterno} the entire analysis in this
paper is classical. We consider a thick SNS weak link driven by a local on-chip microcoil
and ask a single quantitative question: what phase--flux coefficient
$\alpha_{\mathrm{exp}}(d,\omega)$ is actually realised in a realistic device?

The main points can be summarised as follows:
\begin{enumerate}
  \item We focus on thick SNS weak links, a well-known but relatively underused
        regime for magnetic phase control, and show that realistic geometries can
        support phase--flux coefficients much larger than usually assumed for small,
        loopless junctions.

  \item All predictions are derived from the standard Josephson relations, the RCSJ
        model and linear electrodynamics. No nonstandard physics is required: the
        strong-coupling window arises purely from the combination of kinetic
        inductance, mutual inductance and a moderately high-$Q$ plasma mode.

  \item The phase response is characterised by a single dimensionless coefficient
        $\alpha(d,\omega)$ that condenses the coarse-grained GL+Maxwell dynamics of
        the thick SNS region into an effective circuit parameter. In this sense
        $\alpha_{\mathrm{exp}}(d,\omega)$ measures the integrated phase sensitivity of
        a macroscopic mode rather than the properties of individual Andreev states or
        vortices.

  \item If a strong-coupling regime $|\alpha_{\mathrm{exp}}|\gtrsim 0.3$ is realised, thick
        SNS bridges with microcoils effectively become compact loopless phase
        elements: locally addressable ``phase ports'' capable of modulating the
        Josephson phase without a macroscopic flux loop.

  \item Such elements would provide new design options for large-scale
        superconducting circuits, including individually addressable phase or flux
        control of qubits and resonators with reduced routing complexity, and the
        possibility of ultra-compact tunable couplers, on-chip flux pumps,
        travelling-wave phase shifters and building blocks for parametric amplifiers.

  \item Even a negative or intermediate outcome would be informative, by pinning
        down the actual $\alpha_{\mathrm{exp}}(d,\omega)$ realised in thick SNS devices
        and thereby constraining the effective GL+Maxwell descriptions used to model
        them.
\end{enumerate}

In short, the proposed experiment is not a test of any exotic theoretical construct,
but a quantitative probe of how efficiently a realistic thick SNS weak link converts
local microwave magnetic fields into Josephson phase, within the well-established
framework of classical Josephson physics.

%======================================================================
\section{Conclusion}
%======================================================================

We have presented a purely classical analysis of magnetic phase control in thick SNS
weak links, together with a concrete experimental protocol based on RF- and
magnetically driven Shapiro steps. Within a standard RCSJ plus circuit description we
find that realistic devices can in principle reach a strong-coupling regime in which a
local on-chip microcoil modulates the Josephson phase with an efficiency of up to
30--60\% of an ideal dc SQUID.

The proposed experiment extracts an operational coefficient
$\alpha_{\mathrm{exp}}(d,\omega)$ by directly comparing RF-calibrated and
microcoil-driven Shapiro steps. A value $|\alpha_{\mathrm{exp}}|\ll 10^{-2}$ would confirm
the conventional view that local magnetic control of phase in small, loopless junctions
is negligible. A value in the range $|\alpha_{\mathrm{exp}}|\sim 0.3$--$0.6$ would instead
demonstrate unexpectedly strong magnetic-to-phase conversion in thick SNS weak links
and establish them as practical loopless phase elements for superconducting circuitry.
Intermediate values would pinpoint which aspects of the classical modelling --- kinetic
inductance, mutual inductance, quality factor, field distribution --- require refinement.

Either way, the outcome will clarify the role of thick SNS weak links in future
superconducting devices and provide concrete guidance for designing circuits with dense,
locally addressable phase control.

\newpage
%======================================================================
\section*{Appendix A: Illustrative phase-field simulation of the effective mode}
%======================================================================

To illustrate the physical picture behind the coarse-grained description used in the
main text, we performed a minimal one-dimensional phase-field simulation of a thick
SNS bridge under local microwave drive. The model is intentionally simplified and
serves only as a qualitative visualisation; it is not intended as a quantitative
device-level calculation.

The Josephson phase $\phi(x,t)$ along the weak link (in the $x$-direction) is assumed
to obey a damped driven sine-Gordon-type equation
\begin{equation}
  \frac{\partial^2 \phi}{\partial t^2}
  + \gamma \frac{\partial \phi}{\partial t}
  - c^2 \frac{\partial^2 \phi}{\partial x^2}
  + \sin\phi = f(x,t),
  \label{eq:appendix_sG}
\end{equation}
where $\gamma$ is a phenomenological damping coefficient, $c$ is a normalised
Swihart-like velocity, and the local AC drive from the on-chip microcoil is modelled as
\begin{equation}
  f(x,t) = A_{\mathrm{drive}}\,p(x)\cos(\omega t),
  \label{eq:appendix_drive}
\end{equation}
with $p(x)$ a Gaussian profile centred on the bridge,
\begin{equation}
  p(x) = \exp\!\left[-\frac{(x-x_0)^2}{2\sigma^2}\right],
  \qquad x_0 \approx L/2 .
  \label{eq:appendix_px}
\end{equation}

\begin{figure}[H]
  \centering
  \begin{subfigure}[b]{0.8\linewidth}
    \centering
    \includegraphics[width=\linewidth]{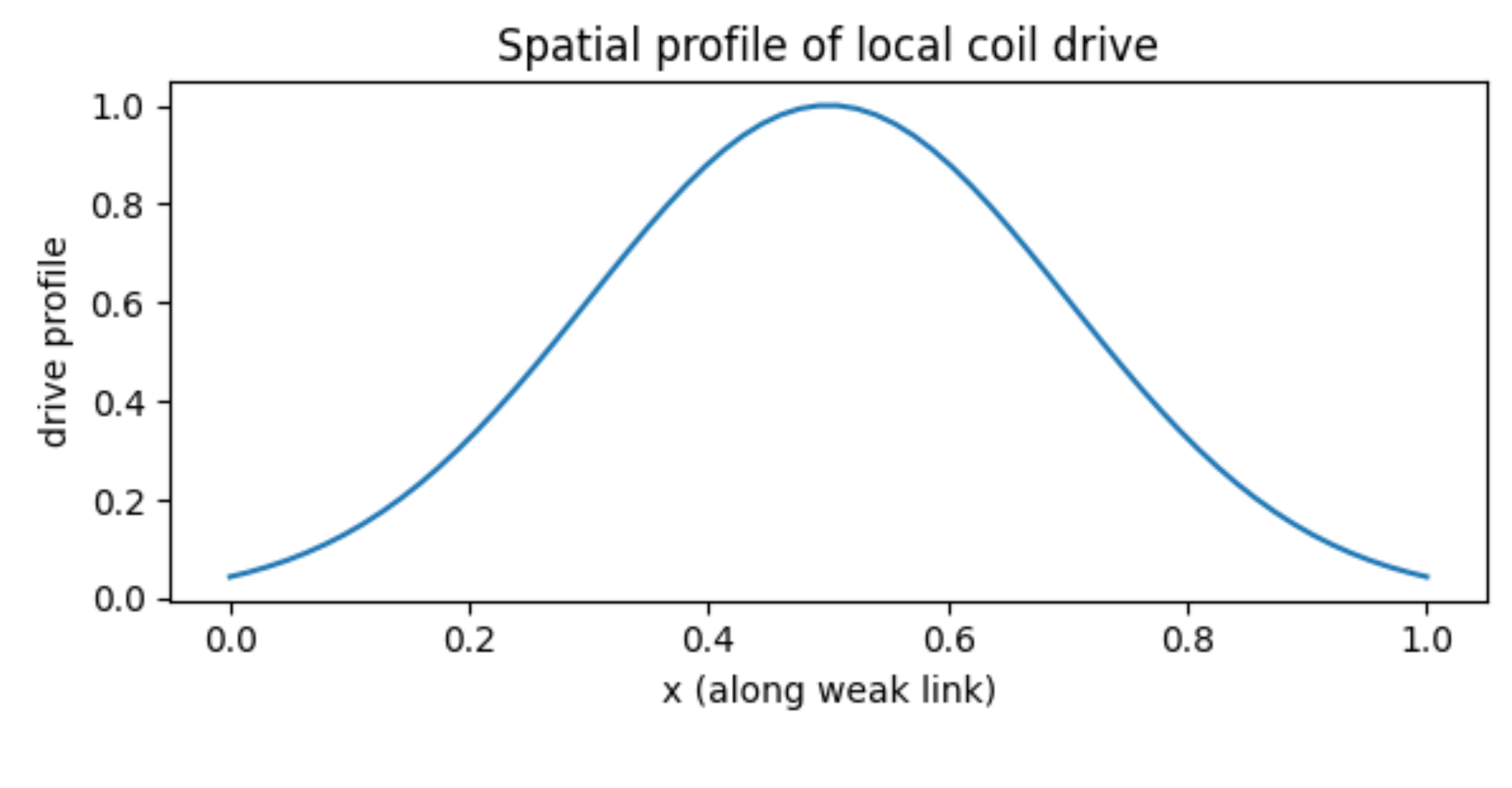}
    \caption{Spatial profile of the local coil drive $p(x)$.}
  \end{subfigure}

  \vspace{1em}

  \begin{subfigure}[b]{0.8\linewidth}
    \centering
    \includegraphics[width=\linewidth]{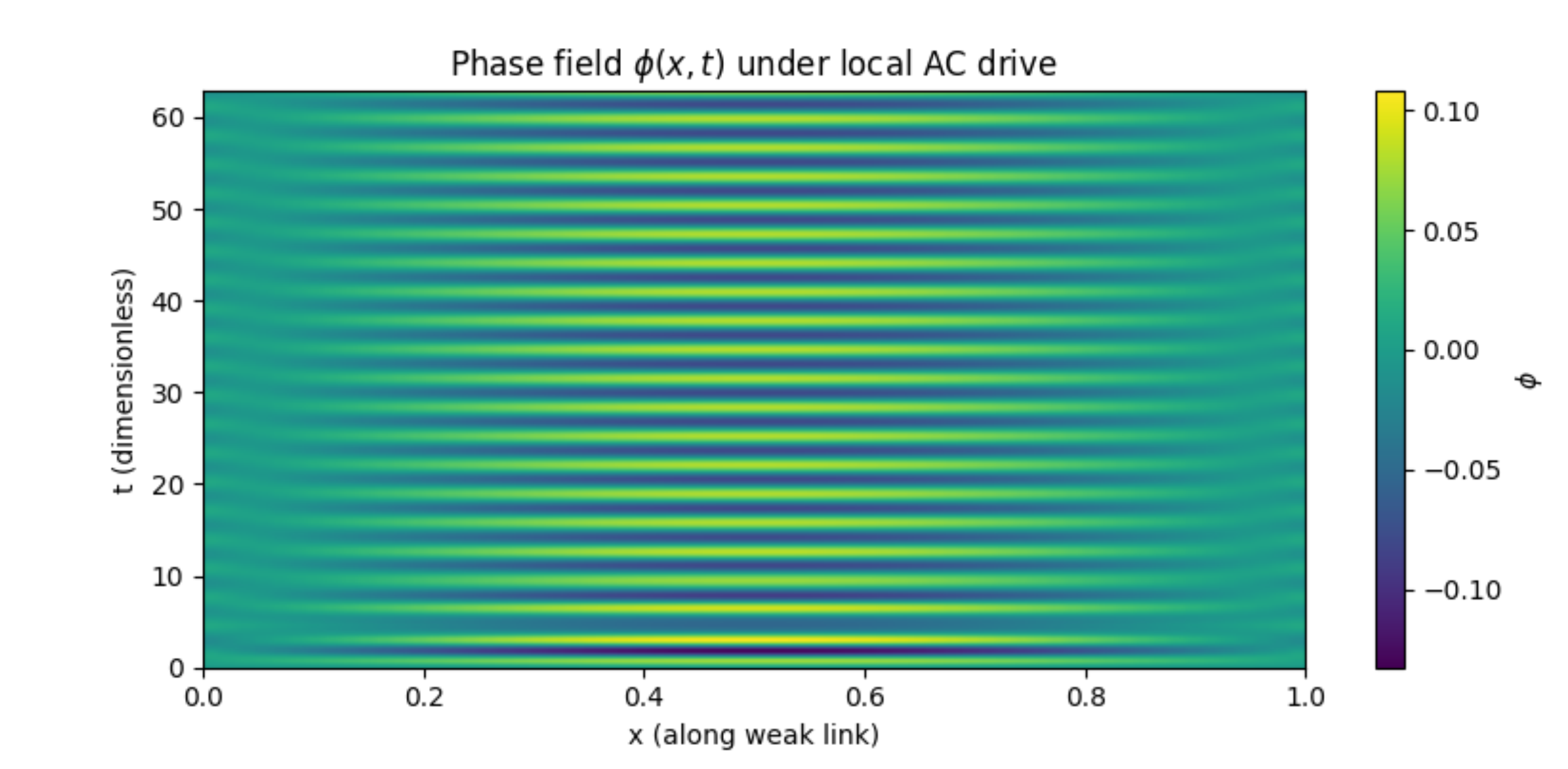}
    \caption{Space--time map of the phase field $\phi(x,t)$ under local AC drive.}
  \end{subfigure}

  \vspace{1em}

  \begin{subfigure}[b]{0.8\linewidth}
    \centering
    \includegraphics[width=\linewidth]{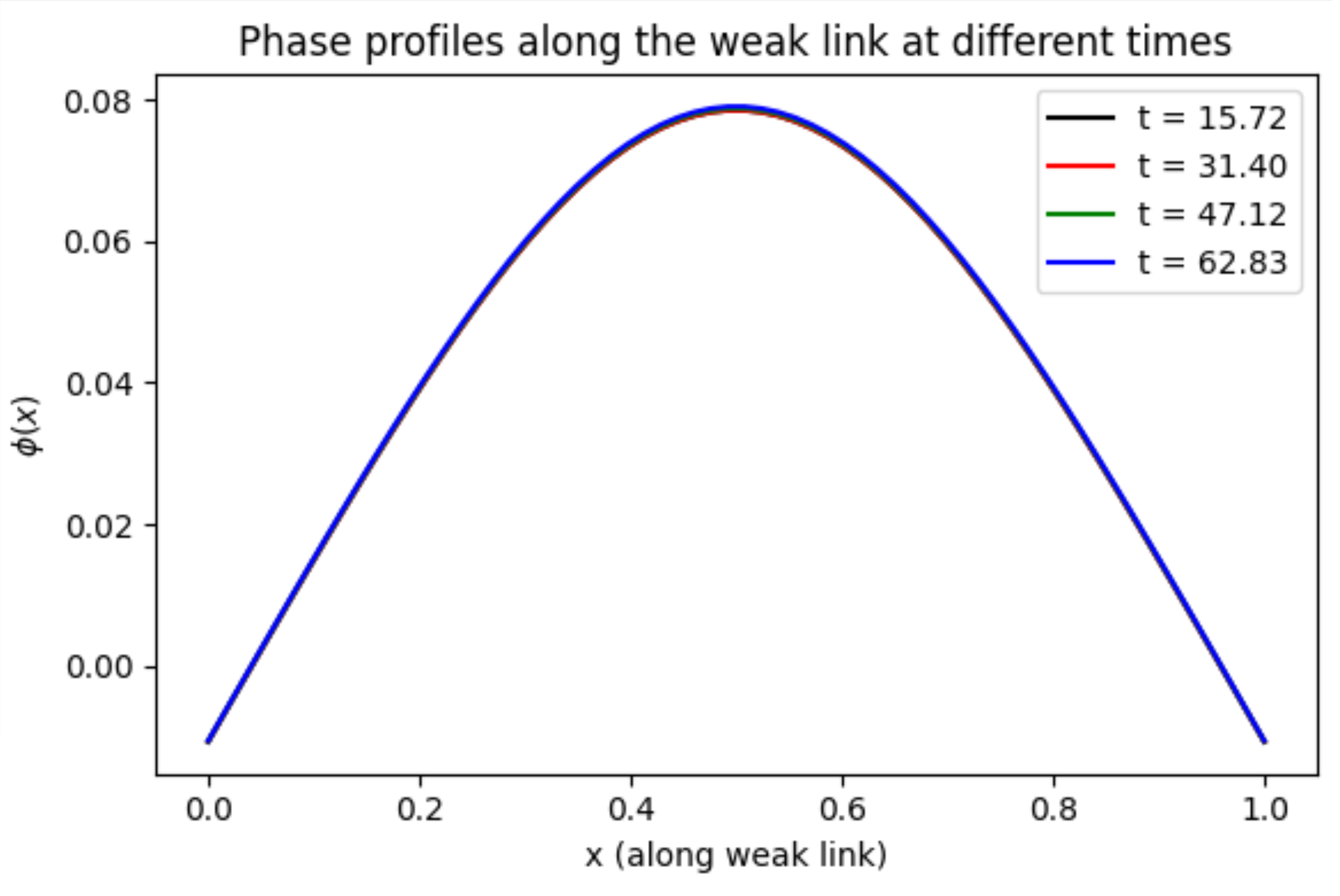}
    \caption{Phase profiles $\phi(x)$ at several times, showing a stable spatial mode.}
  \end{subfigure}

  \caption{Illustrative phase-field simulation for a thick SNS weak link driven by a
  local microcoil. The panels demonstrate that a localised drive predominantly excites
  a single standing-wave-like effective phase mode confined to the central part of the
  bridge.}
  \label{fig:appendix_A1}
\end{figure}

Figure~\ref{fig:appendix_A1} illustrates the resulting phase dynamics for a
representative choice of dimensionless parameters. Panel~(a) shows the spatial
distribution of the drive $p(x)$. Panel~(b) presents a space--time colour map of
$\phi(x,t)$, clearly revealing that the drive predominantly excites a single
standing-wave-like effective mode confined to the central region of the weak link.
Panel~(c) displays instantaneous phase profiles at different times, confirming that the
spatial shape of the excited mode remains essentially unchanged while its amplitude
oscillates.

\begin{figure}[H]
  \centering
  \begin{subfigure}[b]{0.8\linewidth}
    \centering
    \includegraphics[width=\linewidth]{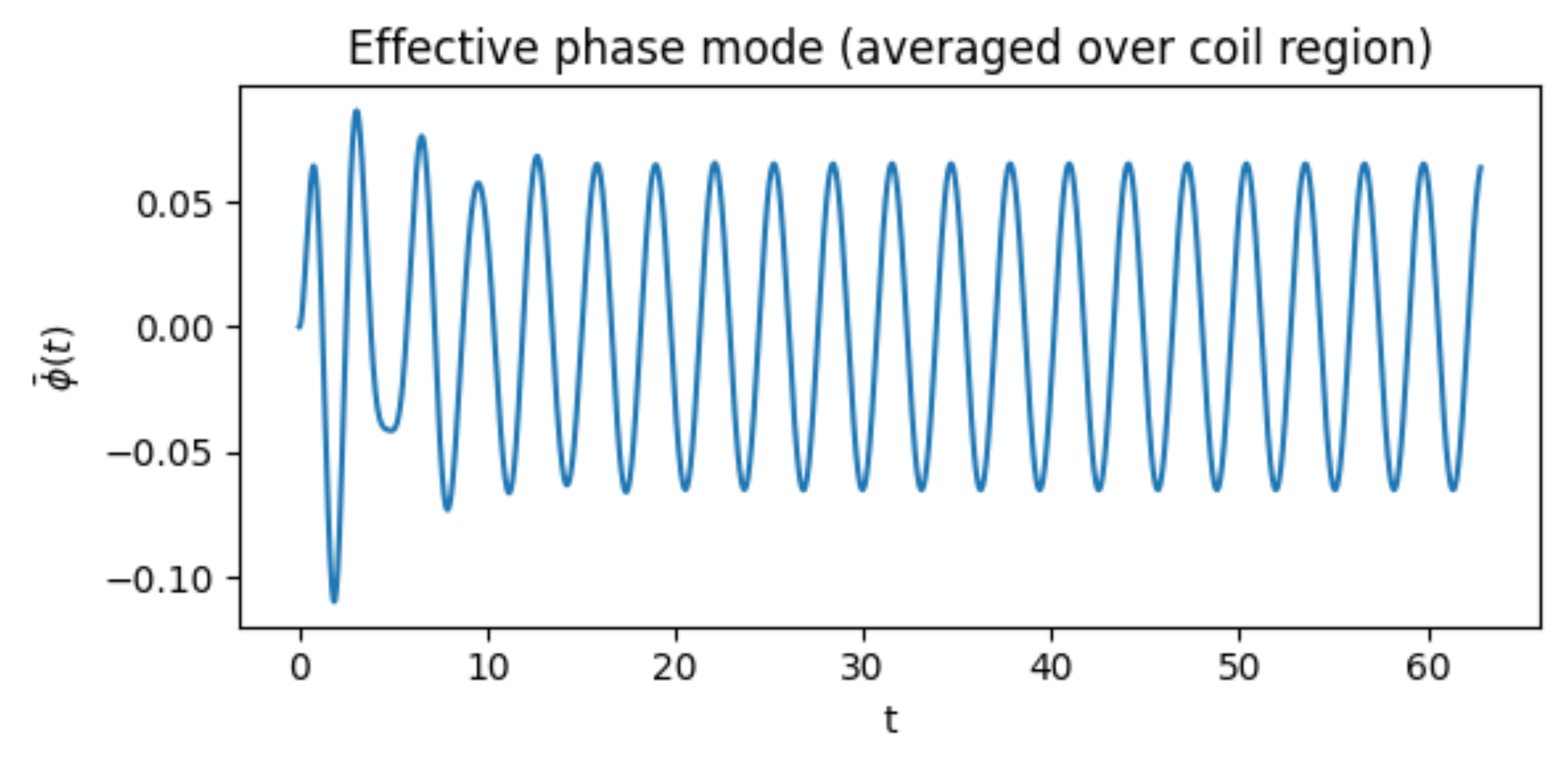}
    \caption{Effective phase mode $\bar{\phi}(t)$ averaged over the coil region.}
  \end{subfigure}

  \vspace{1em}

  \begin{subfigure}[b]{0.8\linewidth}
    \centering
    \includegraphics[width=\linewidth]{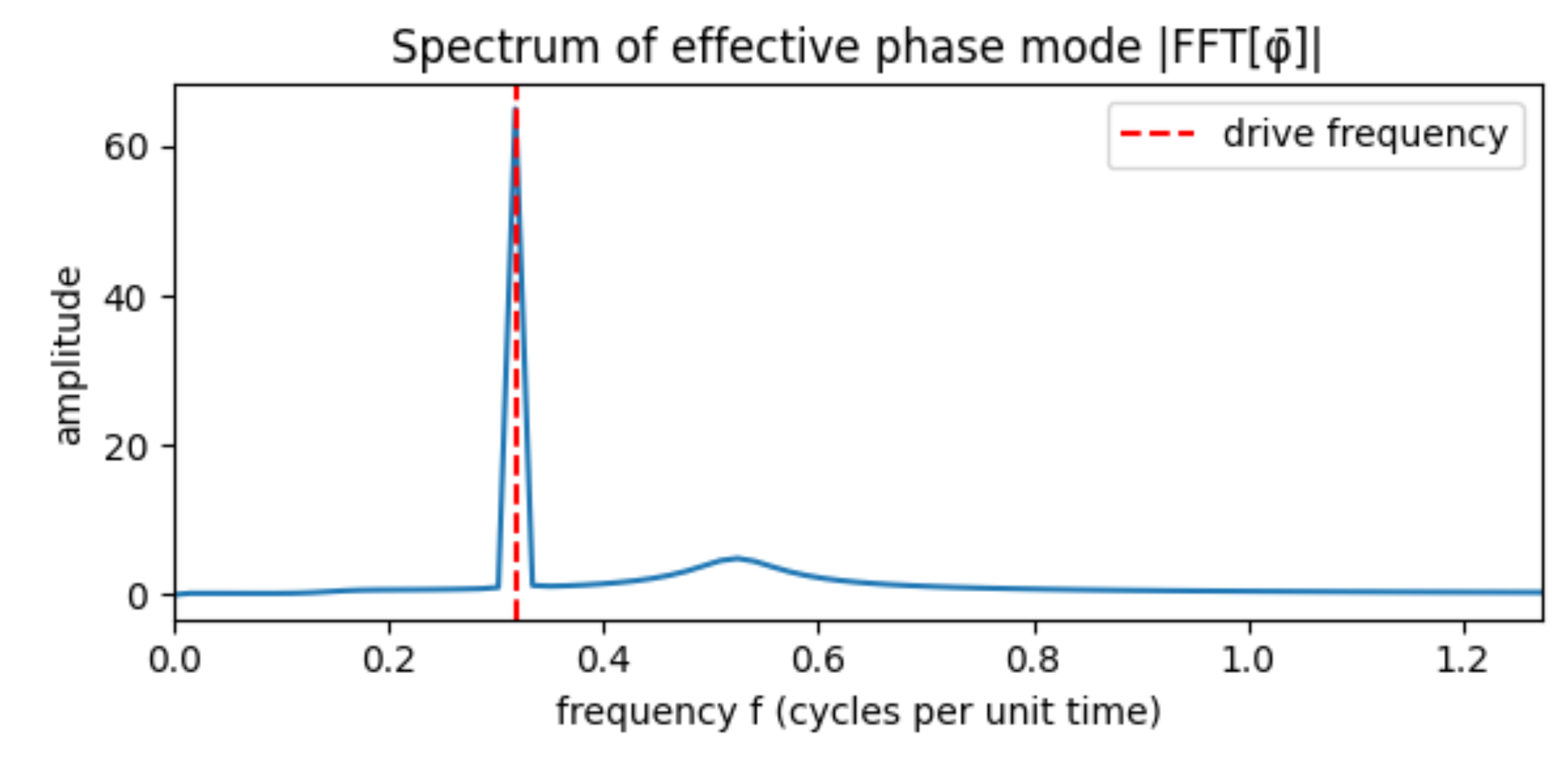}
    \caption{Spectrum $|\mathrm{FFT}[\bar{\phi}]|$ with a peak at the drive
    frequency.}
  \end{subfigure}

  \caption{Time-domain and spectral view of the effective phase mode extracted from the
  phase-field simulation. The sharp spectral peak at the drive frequency illustrates
  the resonant character of the response, consistent with the enhancement of
  $|\alpha(d,\omega)|$ discussed in the main text.}
  \label{fig:appendix_A2}
\end{figure}

To make direct contact with the effective single-degree-of-freedom description in the
main text, we define a spatially averaged phase
\begin{equation}
  \bar{\phi}(t) \equiv
  \frac{\displaystyle\int_0^L \phi(x,t)\,p(x)\,dx}
       {\displaystyle\int_0^L p(x)\,dx},
  \label{eq:appendix_phi_bar}
\end{equation}
weighted by the coil-field profile itself (other reasonable weightings yield virtually
identical results). Figure~\ref{fig:appendix_A2}(a) shows the resulting time trace
$\bar{\phi}(t)$, which rapidly reaches a steady-state sinusoidal oscillation after a short
transient. The Fourier spectrum in Fig.~\ref{fig:appendix_A2}(b) exhibits a sharp peak
at the drive frequency, with amplitude enhanced by roughly the quality factor of the
mode --- behaviour consistent with the resonant enhancement of
$|\alpha(d,\omega)|$ discussed in the main text.

Although all parameters and units are dimensionless and chosen for numerical
convenience, this minimal model illustrates the key physical point: a highly localised
microwave magnetic field excites one dominant effective phase mode in a thick SNS weak
link, whose amplitude shows the same frequency-dependent resonant enhancement that
underlies the strong-coupling regime with $|\alpha|\gtrsim 0.3$ predicted classically in
this work. In particular, the narrow linewidth of the spectral peak in
Fig.~\ref{fig:appendix_A2}(b) corresponds to an effective quality factor
$Q_{\mathrm{eff}}$ of order $10$--$30$, comparable to the values used in the classical
phase--flux model of Sec.~\ref{sec:theory}.

\newpage
%======================================================================
% Bibliography -- copy your existing entries here
%======================================================================

\end{document}